\documentclass[aps,amssymb,showpacs,twocolumn]{revtex4-1}

\usepackage{graphicx}
\usepackage{hyperref}
\usepackage{color}
\usepackage{setspace}

\begin{document}
%\doublespacing
\title{Anomalous Hall Effect in perpendicularly magnetized Mn$_{3-x}$Ga thin films}

\author{M.\ Glas}
\email{mglas@uni-bielefeld.de}
\homepage{www.spinelectronics.de}
\affiliation{Thin Films and Physics of Nanostructures, Bielefeld University, Germany}
\author{D.\ Ebke}
\affiliation{Thin Films and Physics of Nanostructures, Bielefeld University, Germany}
\affiliation{Max Planck Institute for Chemical Physics of Solids, Dresden, Germany}
\author{I.-M.\ Imort}
\affiliation{Thin Films and Physics of Nanostructures, Bielefeld University, Germany}
\author{P.\ Thomas}
\affiliation{Thin Films and Physics of Nanostructures, Bielefeld University, Germany}
\author{G.\ Reiss}
\affiliation{Thin Films and Physics of Nanostructures, Bielefeld University, Germany}

\begin{abstract}
Mn$_{3-x}$Ga (x\,=\,0.1, 0.4, 0.7) thin films on MgO and SrTiO$_3$ substrates were investigated with magnetic anisotropy perpendicular to the film plane. An anomalous Hall-effect was observed for the tetragonal distorted lattice in the crystallographic D0$_{22}$ phase. The Hall resistivity $\varrho_{xy}$ was measured in a temperature range from 20 to 330\,K. The determined skew scattering and side jump coefficients are discussed with regard to the film composition and used substrate and compared to the crystallographic and magnetic properties.

\end{abstract}

\maketitle

Recently, the integration of materials with perpendicular magnetic anisotropy into magnetic tunnel junctions (MTJs) have found a lot of attraction due to the predicted lowered current densities for current induced switching and a higher thermal stability.\cite{Kent:2004p3669}
The perpendicular anisotropy of magnetic thin films for the integration into MTJs can be realized differently.

One example for common perpendicular thin films are $(\rm{Co/Pd})_n$\cite{Kugler:2011p3334} or $(\rm{Co/Pt})_n$\cite{Carcia:1985p2264} multilayers which take advantage of interface anisotropy.

Another approach to induce a perpendicular magnetic anisotropy is the utilization of ferromagnetic materials in combination with rare earth elements. The integration of such films into so-called perpendicular MTJs (pMTJs) or spin valves (SPVs) was reported by Nakayama et al.\cite{Nakayama:2008p3670} and Li et al.\cite{Li:2010p3257} for Tb+CoFe and Tb+Co$_2$FeAl.

However, for the industrial applicability a single perpendicular film is preferred to avoid process complications and to realize tunnel junctions and spin valves with a total thickness below 10\,nm.

Furthermore, materials with a high spin polarization like Heusler compounds \cite{DeGroot:1983p65} are eligible for the integration into MTJs to realize high tunneling magnetoresistance (TMR) ratios. Such junctions with half-metallic Heusler junctions have been successfully demonstrated in the recent years by several groups for in-plane magnetization.\cite{Ebke:2010p1637,Wang:2009p858,Tsunegi:2008p68,Tezuka:2009p912}

A high spin polarization and an out-of-plane magnetization is predicted for the Heusler compound Mn$_3$Ga.\cite{Balke:2007p2533,Winterlik:2008p2531,Wurmehl:2006p2224}
Such films exhibit a large perpendicular magnetic anisotropy and experimental values of up to $1.2 \times 10^7$\,erg/cm$^3$ are reported by Wu et al. for the effective anisotropy constant $K_u^{eff}$.\cite{Wu:2010p4215}
Furthermore, the spin polarization of Mn$_3$Ga thin films was determined experimentally by point-contact Andreev reflection (PCAR). Here, a maximum value of 58\% is reported by Kurt et al.\cite{Kurt:2011p3666}
Additionally, a high spin polarization of Mn$_2$Ga is concluded by Wu et al. from electrical transport properties studies using anomalous (or extraordinary) Hall-effect.\cite{Wu:2010p4215}
Therefore, the Mn$_{3-x}$Ga compounds fulfill all required items of a high performance spin torque device suggested by the shopping list of Graf et al.\cite{Graf:2011jj}

%But the influence of stoichiometry in Mn$_{3-x}$Ga is still unclear.

In this work, we have investigated the magnetic, crystallographic and transport properties of Mn$_{3-x}$Ga thin films (x\,=\,0.1, 0.4, 0.7). A magnetic perpendicular anisotropy of these layers can be realized when grown in the crystallographic D0$_{22}$ phase (Figure 1a)).

DC and RF magnetron sputtering was used for the preparation of our thin films. The Mn$_{3-x}$Ga layers were co-deposited in a UHV system from a Mn$_{50}$Ga$_{50}$ and a Mn sputtering target to adjust the stoichiometry which was verified by X-ray fluorescence (XRF) measurements.
The Mn-Ga layers were deposited on MgO (001) and SrTiO$_3$ (001) (STO) substrates at different substrate temperatures of up to 650$^\circ$C to generate a tetragonal distorted crystal structure. A layer thickness of $23$, $25$ and $27\,$nm was determined for the Mn$_{2.3}$Ga, Mn$_{2.6}$Ga and Mn$_{2.9}$Ga films.
The Ar pressure was $1.3 \times 10^{-3}$ mbar and the growth rate was set to $\approx$ 0.4\,nm/sec.
The lattice mismatch of $7\,\%$ for MgO (a$_{MgO}=4.21$\,\AA) and $<1\,\%$ for SrTiO$_3$ (a$_{STO}=3.91$\,\AA) allows a coherent growth of the Mn$_{3-x}$Ga films (a$_{MnGa}\approx3.9$\,\AA) in c axis orientation. Finally, the layers were capped by a 2\,nm thick MgO layer to prevent the Mn$_{3-x}$Ga from oxidation. 

\begin{figure}[b]
\includegraphics[width=\linewidth]{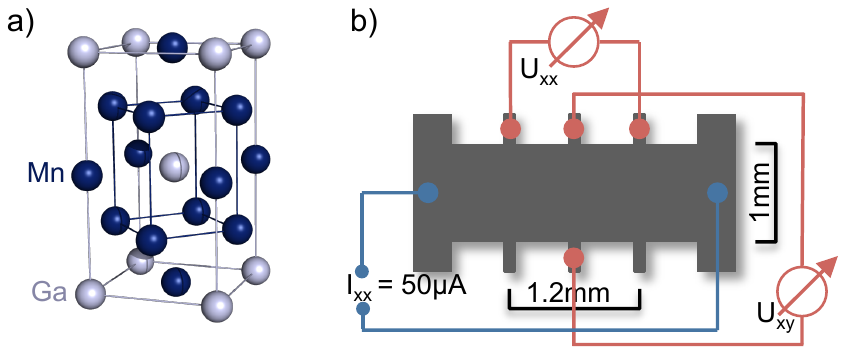}
\caption{a) Crystal structure of the tetragonal Mn$_3$Ga in D0$_{22}$ phase. In plane lattices $a,b\approx3.9$\,\AA; out of plane lattice $c\approx7.2$\,\AA. b) Schematically structure of the used Hall bar and measurement geometry.}
\label{fig:Fig1}
\end{figure}

Figure 1b) schematically shows the used Hall bars ($1$\,mm wide $\times$ $2.5$\,mm long), which were patterned by optical lithography and ion beam etching. Contact pads were fabricated by depositing Ta ($5$\,nm) and Au ($40$\,nm). 
The measurements of the transport properties were carried out by 4-terminal arrangement in a temperature range between $20$\,K and $330$\,K. The Hall voltage was measured in a magnetic field of up to $1$\,T perpendicular to the film plane.

X-ray diffraction (XRD) studies were performed by a Philips X'pert pro diffractometer ($\lambda_{Cu,k_\alpha}=1.5418\,$\AA) to find the optimal deposition temperature for the Mn$_{3-x}$Ga layers. The intended tetragonal D0$_{22}$ phase was found for deposition temperatures from 490$^{\circ}$C to 570$^{\circ}$C. The diffraction pattern of the Mn-Ga films which were prepared at lower deposition temperatures show peaks which can be attributed to the cubic D0$_{3}$ phase. Higher deposition temperatures leads to a phase change into the hexagonal D0$_{19}$ phase.

\begin{figure}[t]
\includegraphics[width=\linewidth]{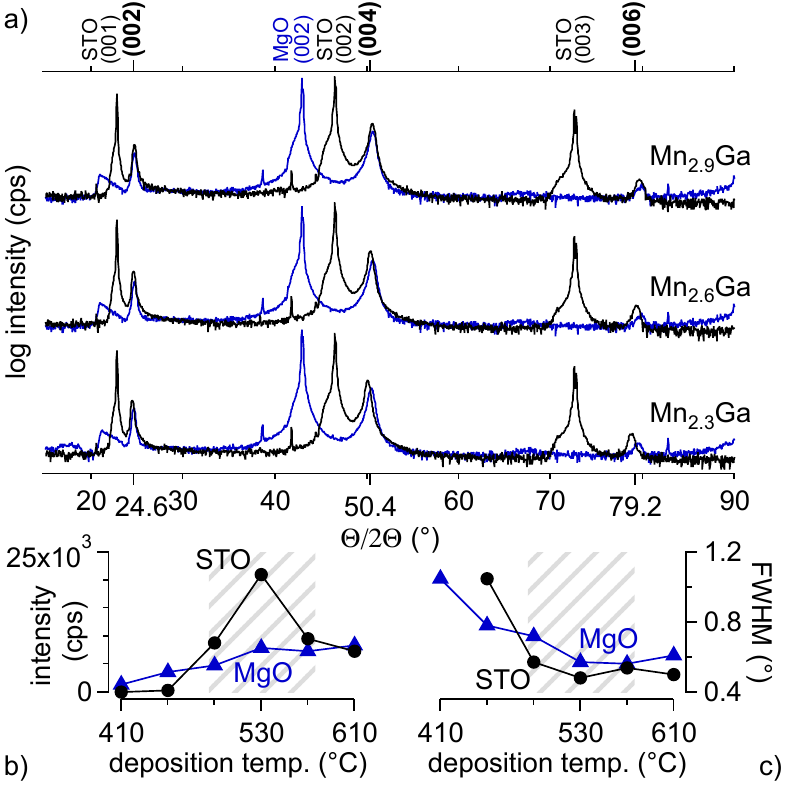}
\caption{a) XRD pattern of 25\,nm thick Mn-Ga layers with varying stoichiometry deposited on MgO (blue) and STO (black) substrates at 530$^{\circ}$C. Mn$_{2.9}$Ga (004) superlattice peak intensity b) and FWHM c) as a function of deposition temperature for MgO and STO substrates.}
\label{fig:Fig2}
\end{figure}

Figure 2a) shows the obtained $\Theta$/2$\Theta$ scans for the investigated Mn-Ga compositions deposited at 530$^{\circ}$C on MgO and STO substrates, respectively. The (002) and (004) superlattice peaks are clearly visible at 24.6 and 50.4 degree. Additionally, the presence of the (006) peak at 79.2 degree demonstrates a high film quality. The substrate peaks can be found at 42.9 (002) degree for MgO and at 22.8 (001), 46.5 (002) and 72.6 (003) degree for STO.
The corresponding \(c\) axis lattice constants ranges are $c_{MgO}= 7.18$ to $7.22$\,\AA~ on MgO substrates and $c_{STO}= 7.22$ to $7.25$\,\AA~ on STO substrates. In both cases the lattice expands slightly with decreasing Mn content. Winterlik et al.\ also observed this behavior and attributed it to the removal of Mn of the X positions.\cite{Winterlik:2008p2531} The different \(c\) axis parameter for MgO and STO originate from the lattice mismatch in the \(a\)-\(b\) plane.
The Mn-Ga in-plane lattice constants ($a,b\approx3.9$\,\AA) perfectly matches the STO substrate lattice ($a_{STO}=3.91$\,\AA). This leads to a higher crystallinity of the Mn-Ga films when compared to these grown on MgO substrates. 
As shown in Figure 2b) a higher (004) peak intensity and a lower full width at half maximum (FWHM) (Figure 2c)) was determined from the XRD pattern for the Mn$_{2.9}$Ga films which were deposited on STO. Here, a minimum FWHM of 0.48 degree was achieved for a deposition temperature of 530$^{\circ}$C. The values are as low as the values reported by Wu et al. \cite{Wu:2009p2744} for similar films and indicate the good crystalline quality of these films.
The intended tetragonal D0$_{22}$ phase of the Mn-Ga films was verified by pole figure scans. The fourfold (011) peak could be observed for all investigated stoichiometries in a deposition temperature range from 490$^{\circ}$C to 570$^{\circ}$C. 

\begin{figure}[t!]
\includegraphics[width=\linewidth]{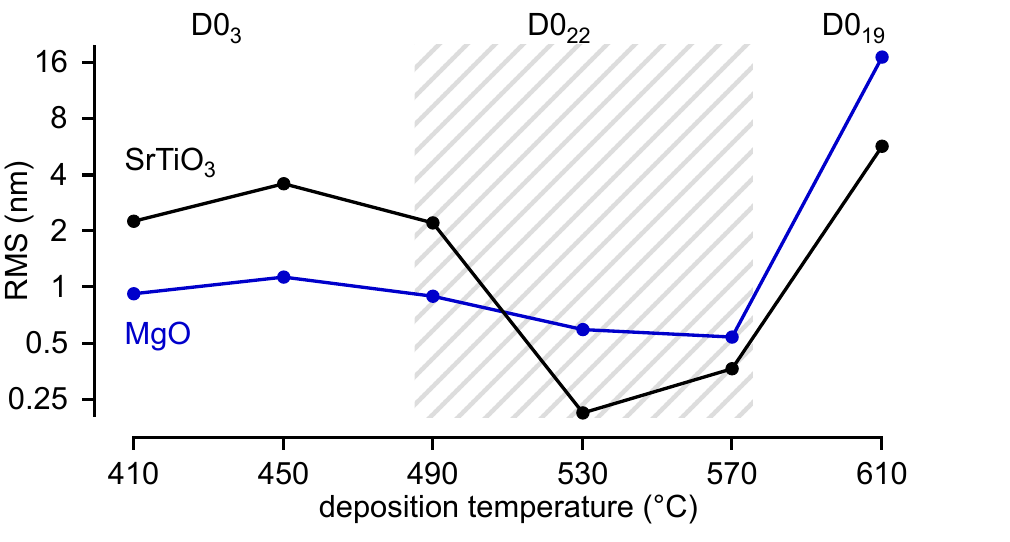}
\caption{root mean square (RMS) as function of the deposition temperature for Mn\(_{2.9}\)Ga thin films on MgO (blue) and STO (black) substrates.}
\label{fig:Fig3_1}
\end{figure}

To investigate the surface roughness of the Mn\(_{3-x}\)Ga thin films, atomic force microscopy (AFM) measurements were performed. The surface structure is an important factor for the further integration of Mn\(_{3-x}\)Ga into magnetic tunnel junctions. A high roughness will produce pinholes through the barrier, thus the tunnel probability tends to zero. Figure 3 presents the rms surface roughness in relation to the deposition temperature for a 27\,nm thick Mn\(_{2.9}\)Ga thin film. The rms surface roughness obtained a minimum value in the temperature regime between 490 and 570\(^{\circ}\)C, where the crystallographic properties are ideal. For thin films on STO substrates the proper deposition temperature is 530\(^{\circ}\)C (0.21\,nm rms) and on MgO substrates 570\(^{\circ}\)C (0.54\,nm rms). The films deposited at 530\(^{\circ}\)C on STO show a roughness comparable with the data obtained by Wu et al.\cite{Wu:2010p2745} Our films grown on MgO are smoother than films prepared by Kurt et al. Samples prepared on STO showed a smoother surface when compared to Pt-buffered samples, due to the lower lattice mismatch.\cite{Kurt:2011p3666}

\begin{figure}[t!]
\includegraphics[width=\linewidth]{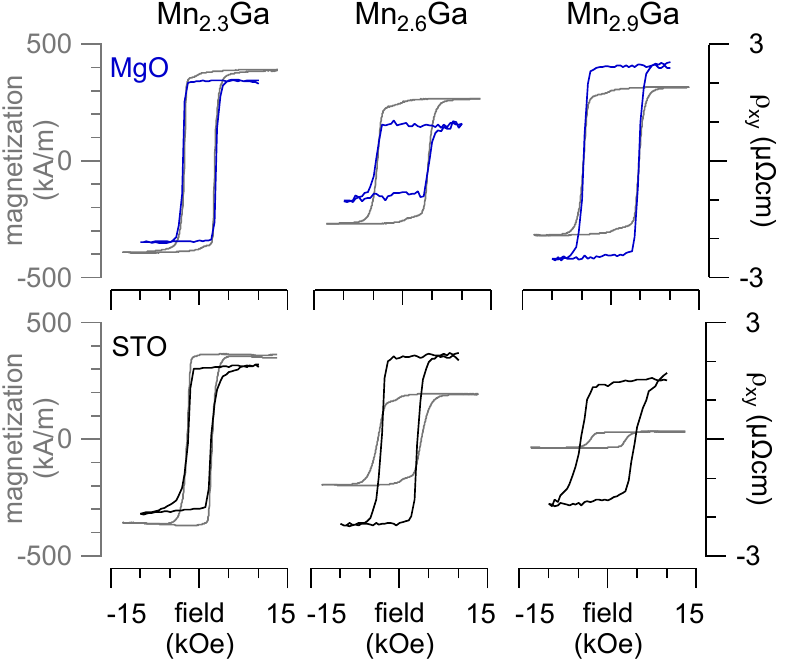}
\caption{Overlay of magnetization data (grey) and transversal Hall resistivity for Mn$_{3-x}$Ga deposited on MgO (top) and STO (bottom) substrates at 530$^{\circ}$C.}
\label{fig:Fig3}
\end{figure}

\begin{figure}[t]
\includegraphics[width=\linewidth]{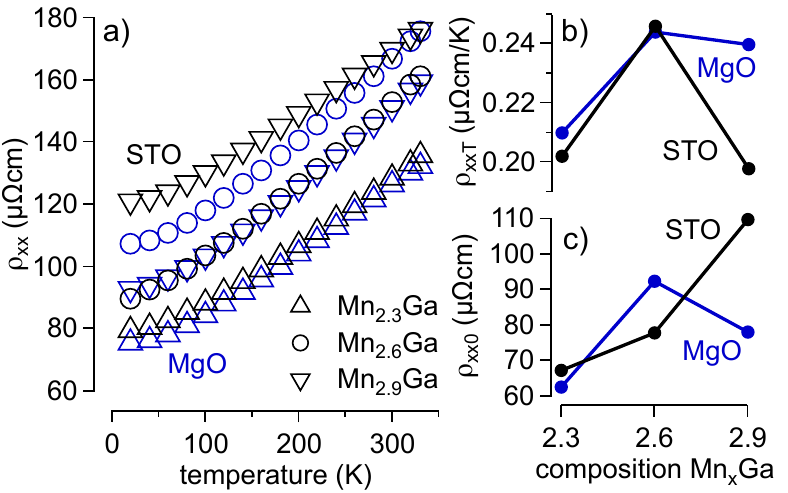}
\caption{a) Temperature dependence of the longitudinal resistivity $\varrho_{xx}$. b) extracted temperature dependent contribution $\varrho_{xxT}$ c) residual resistivity $\varrho_{xx0}$}
\label{fig:Fig4}
\end{figure}

Figure 4 shows an overlay of magnetization data (grey) and the transversal Hall resistivity $\varrho_{xy}$ for all compositions deposited at 530$^{\circ}$C.
A maximum value of $\varrho_{xy}^{sat}=2.2\,\mu\Omega$cm can be found for the saturated Hall resistivity of the Mn$_{2.9}$Ga film deposited on MgO.
All values are lower when compared to Mn$_2$Ga films reported by Wu et al. ($11.5\,\mu\Omega$cm)\cite{Wu:2010p4215} which might originate from different preparation conditions such as growth rate, pressure and temperature.
As expected, we found a decreasing magnetic moment with increasing Mn content for the layers deposited on STO. The decrease in the magnetic moment originates again from a higher removal of Mn on the X positions compared to the Y positions, which leads to a nonequilibrium of compensating Mn moments.\cite{Winterlik:2008p2531}
By contrast, the lowest moment for films deposited on MgO was found for Mn$_{2.6}$Ga. Most likely, the higher magnetization we found for Mn$_{2.9}$Ga is due to atomic disorder. In addition, the magnetic moment for thin films on MgO is higher than for layers on STO substrates. This characteristic arise from the bigger expansion in the \(a\)-\(b\) plane for Mn\(_{3-x}\)Ga on MgO substrates. Due to this expansion the cell area increases with decreasing height, which leads to a weakening in the compensation between the Mn atoms on the X and Y positions.

To quantify the degree of order in our layers, we have analyzed the temperature dependence of the longitudinal resistivity $\varrho_{xx}$ (Figure 5a) which can be expressed as $\varrho_{xx}(T)=\varrho_{xx0}+\varrho_{xxT}(T)$. The values were determined by a linear fit of  $\varrho_{xx}$ in the temperature range of 100\,K to RT. The temperature dependent contribution of the resistivity $\varrho_{xxT}$ stands for the degree of electron scattering on phonons and magnons.
The highest values were observed for Mn$_{2.9}$Ga on MgO and Mn$_{2.6}$Ga (Figure 5b)).
Furthermore, the $\varrho_{xx0}$ represents atomic disorder or impurities (Figure 5c)).
The extracted values scale reverse the magnetic moments.
A maximum $\varrho_{xx0}$ was found for the nearly compensated Mn$_{2.9}$Ga film deposited on STO. Because of the higher film quality we found for layers grown on STO this is most likely due to the antiparrallel aligned Mn moments instead of atomic disorder. 

\begin{figure}[t!]
\includegraphics[width=\linewidth]{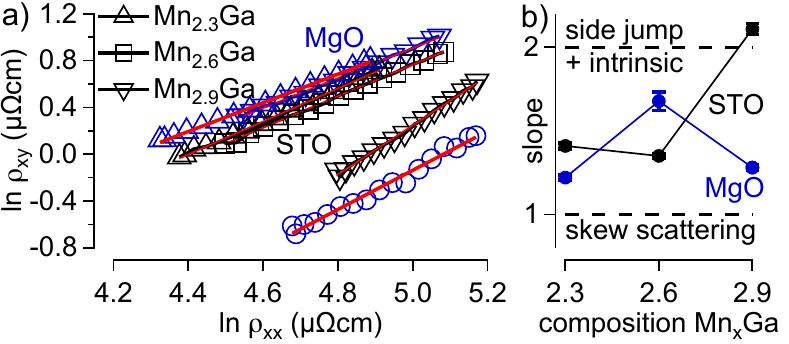}
\caption{a) measured Hall resistivity ploted as ln $\varrho_{xy}$ vs ln $\varrho_{xx}$. b) slope of ln $\varrho_{xy}$ vs ln $\varrho_{xx}$ linear fit.}
\label{fig:Fig5}
\end{figure}

\begin{figure}[t!]
\includegraphics[width=\linewidth]{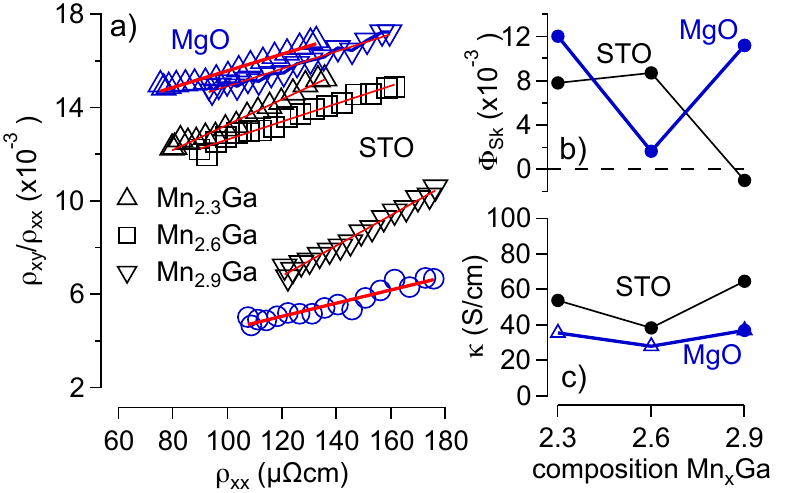}
\caption{a) the line fit of the data ratio $\varrho_{xy}/\varrho_{xx}$ vs $\varrho_{xx}$ to determine the skew scattering parameter $\Phi_{sk}$ b) and $\kappa$ c)}
\label{fig:Fig6}
\end{figure}

Generally, the Hall resistivity consists of two contributions: $\varrho_{xy}=R_0H+R_S4\pi M_S$, where $R_0$ is the ordinary Hall coefficient and $R_S$ the anomalous Hall coefficient.
Because $\varrho_{xy}$ in Figure 4 has a similar shape to the magnetization hysteresis loops, the AHE is much larger than the ordinary Hall effect in the investigated samples. The anomalous Hall resistivity depends on intrinsic and extrinsic effects and can be expressed as: $\varrho_{xy}=\Phi_{sk} \varrho_{xx}+\kappa \varrho_{xx}^2$

Here, the parameter $\Phi_{sk}$ of the linear term of the longitudinal resistivity $\varrho_{xx}$ represents the extrinsic skew scattering.\cite{Smit:1958p4317}
The $\kappa$ of the quadratic term contains the extrinsic conductivity $\kappa_{sj}$, which originates from the side-jump mechanism and the intrinsic conductivity $\kappa_{int}$, which originates from the band structure.\cite{Berger:1964p4319}

A linear fit of the function ln $\varrho_{xy}$ vs ln $\varrho_{xx}$ is applied to distinguish the dominating mechanism from the experimental data. The data and the corresponding fit functions of the investigated samples are given in Figure 6a).
A slope $m$ of the fit function with $m=1$ gives a dominating skew scattering mechanism whereas $m=2$ represents the side jump and the intrinsic mechanism.\cite{Luttinger:1958p3663,Berger:1970p4318}
Except the slope of $m=2.1$ for the Mn$_{2.9}$Ga sample grown on STO, the determined slopes of the fit functions (Figure 6b) indicating a contribution of both mechanism. A dominating skew scattering can be deduced for layers with higher magnetization.

For extended differentiation between skew scattering and intrinsic factor, the values of $\Phi_{sk}$ and $\kappa$ are determined by a linear fit when $\varrho_{xy}/\varrho_{xx}$ is plotted vs $\varrho_{xx}$ as shown in Figure 7a). The resulting slope of the fit characterizes $\kappa$ (Figure 6c)), the y-axis intercept represents $\Phi_{sk}$ (Figure 7b)).
The extracted $\Phi_{sk}$ value for the Mn$_{2.9}$Ga film on STO nearly vanishes ($-1.0 \times 10^{-3}$) as expected from Figure 5b). Additionally, a similar low value ($1.6 \times 10^{-3}$) can be found for Mn$_{2.6}$Ga on MgO. 
The corresponding $\kappa$ values range between $25$ and $64\,$S/cm. Notably, the $\kappa$ values of films grown on STO are throughout higher when compared to these grown on MgO. Probably, this represents the higher ordering we found for STO from XRD measurements.

In summary, we have fabricated Mn$_{3-x}$Ga thin films at various deposition temperatures. Because of the difference in the magnetic moment for different compositions. We conclude that Mn$_{2.9}$Ga is the favorable composition for spintronic devices, because of the low magnetic moment.
We found STO substrates to be preferable when compared with MgO substrates with regard to the crystalline quality of the films. The observed higher $\kappa$ contribution of this layers might be attributed to the ordering. 
Finally, we conclude that the obtained skew scattering is correlated to the magnetic moment of the Mn$_{3-x}$Ga films.

%\newpage

\end{document}